\documentclass{article}
\usepackage[utf8]{inputenc}
\usepackage{graphics,amsmath,amssymb,epsfig,epstopdf}

\title{Analytical reconstruction of equivalent purely kinetic k-essence description for $c_s^2(w)$ barotropic fluid models }

\author{Dalibor Perkovi\'{c}$^{1,}$\thanks{dalibor.perkovic@zvu.hr}  and Hrvoje \v Stefan\v ci\'c$^{2,}$\thanks{hrvoje.stefancic@unicath.hr} }

\date{
\centering
$^{1}$ University of Applied Health Sciences, Mlinarska street 38, 10000 Zagreb, Croatia \\
\vspace{0.2cm}
$^{2}$ Catholic University of Croatia, Ilica 244, 10000 Zagreb, Croatia }

\begin{document}

\maketitle

\abstract{A novel approach to the class of cosmic barotropic fluids in which the speed of sound squared is defined as a function of the Equation of State parameter, so called $c_s^2(w)$ models, is examined. For this class of models, a new analytical reconstruction method is introduced for finding their equivalent purely kinetic k-essence formulation. The method is explicitly demonstrated for several $c_s^2(w)$ models. {\bf A distinguished feature of the studied models is a peak in $c_s^2$ during cosmic expansion. A possible role that such model features could play in understanding of so far unexplained aspects of cosmic evolution is discussed.} The application of the obtained explicit or closed form solutions in understanding dark sector unification models is {\bf outlined}.   }

\section{Introduction}

One of the most dramatic shifts in physical understanding of the universe happened with the discovery of its late-time accelerated expansion. Despite the mounting observational evidence of the speed-up in the global cosmic dynamics \cite{SNIa1,SNIa2,CMB1,CMB2,BAO}, the question of precise mechanism causing the cosmic acceleration remains unanswered. Among numerous proposals for the said mechanism, two of the most prominent approaches are the presence of a cosmic component with negative pressure, called {\em dark energy} \cite{DERev1,DERev2,DERev3,DERev4,DERev5,DEparam}, and modifications of gravitational interaction at cosmic distances \cite{ModGravRev1,ModGravRev2,ModGravRev3,ModGravRev4}. 

An older, though not less enigmatic, challenge is evident in increased gravitational interaction without a visible source at various scales in galactic and galaxy cluster dynamics, as well as in the cosmic history of growth of inhomogeneities and large scale structure formation. The most studied approach is the presence of a cosmic component, called {\em dark matter} \cite{DM1,DM2,DM3,DM4}, although alternatives exist, with MOND and its generalizations being the most notable one \cite{MOND1,MOND2,MOND3,MOND4,MOND5}.   



A cosmic model assuming that dark energy is a small positive cosmological constant and that dark matter comes in the form of cold dark matter, known as the $\Lambda$CDM model, has occupied a central place in cosmology for more than two decades owing to its simplicity and explanatory power. However, ever more precise and more abundant cosmic observations have revealed inadequacies of this model, e.g. in the form of so called $H_0$ tension \cite{H0tension1,H0tension2,H0tension3,H0tension4,H0tension5}. Recent years have brought a large number of various attempts beyond the $\Lambda$CDM model, questioning and extending all its assumptions. 

An idea of a single cosmic component unifying the concepts of dark matter and dark energy is attractive, both for fundamental and practical reasons, see e.g. \cite{Bertacca}. A further assumption that this cosmic component unifying the dark sector can be described as a barotropic perfect fluid \cite{LinderScherrer} comes naturally, since other cosmic components, such as non-relativistic matter or radiation, allow such a description. One of such early and prominent examples of barotropic fluid unifications is the Chaplygin gas \cite{Chap1,Chap2} and its numerous generalizations \cite{GenChap1,ModChap1,ModChap2,HybridChap,Umami} (see also \cite{GCGconstraint} for early strong observational bounds).

Recent works \cite{underdetermined} also claim that the concept of dark energy is underdetermined, i.e. that available cosmological observations cannot fully constrain microscopic dark energy models. The best description of dark energy, then, is of phenomenological character. In view of such claims (but also as a generally sound phenomenological approach), it is important to approach dark sector model building in a way that is consistent with observational data. Very rough and somewhat simplified categorization of observational signals in cosmology could be to those that inform us on the cosmic global expansion and those that provide insight into structure formation and growth. For a dark sector unified component being a barotropic perfect fluid, the global expansion of the universe is influenced by its Equation of State parameter $w \equiv p/\rho$, whereas the quantity of importance with respect to structure growth is its adiabatic speed of sound $c_s^2 \equiv \frac{d p}{d \rho}$. Therefore, it seems phenomenologically advantageous to develop models where these two quantities are related, i.e. where $c_s^2$ is a function of $w$. The usefulness of these models, called $c_s^2(w)$ models, has already been demonstrated in the study of dark sector unifications \cite{Caplar,mi2,mi4}, cosmological constant barrier crossing \cite{mi1}, physical viability of dark energy parametrizations \cite{mi3} and analysis of galaxy rotation curves in models where dark matter is a barotropic fluid \cite{mi5}.  

A powerful field-theoretical framework for the description of the dark sector and its dynamics, based on the generalized kinetic term of the scalar field, is known as {\em k-essence}. K-essence was successfully applied in the description of inflation \cite{CAP1,Garriga}, dark energy \cite{Chiba,CAP2,CAP3,Malquarti1,Malquarti2,Chimento,dePutter}, unified dark sector \cite{Scherrer,Chimento2,Armendariz,kessence2,Bertacca2,kessence3,Bertacca3,Chavanis}, tachyons \cite{tachyon1,tachyon2,tachyon3,tachyon4} {\bf and alleviation of cosmic tensions \cite{kesstension}}. An equivalent microscopic description of the $c_s^2(w)$ models can be obtained in the framework of purely kinetic k-essence \cite{Diez,Arroja,Ferreira,Unnikrishnan,kessphase}. A detailed presentation of a general approach to how to translate  to purely kinetic k-essence models was given in \cite{mi4}, together with a concrete analytic demonstration of the approach for a number of $c_s^2(w)$ models. In this paper we introduce a novel method for the reconstruction of purely kinetic k-essence for the known $c_s^2(w)$ models. The presentation of the general method is followed by its application to five $c_s^2(w)$ models. {\bf The term "reconstruction" is sometimes used for the method(s) of determining the underlying theoretical formulation or model from the observational data.  It is important to stress that in this paper the term "reconstruction" is understood as a method of obtaining an equivalent theoretical formulation of a model (e.g. purely kinetic k-essence) from another theoretical formulation (e. g. $c_s^2(w)$ model).}



The structure of this paper is the following. The first section is the introduction. The method of reconstructing of equivalent purely kinetic k-essence is presented in the second section. The third section brings the application of the method developed in the second section to five distinct $c_s^2(w)$ models. The fourth section closes the paper with the discussion and conclusions.

\section{A new reconstruction method}

In purely kinetic k-essence models the Lagrangian density is dependent on its kinetic term only, i.e. ${\cal L}=F(X)$, where $X=\frac{1}{2} g^{\mu\nu} \nabla_\mu \phi \nabla_\nu \phi$. These models have an equivalent perfect fluid representation \cite{Diez,Arroja,Ferreira}, where the fluid four-velocity is $u_\mu = \frac{\nabla_\mu \phi}{\sqrt{2 X}}$. Furthermore, the energy density has the form
\begin{equation}
\label{eq:kessden}
\rho=2 X F_{X}-F \, ,
\end{equation}
where the subscript $X$ refers to differentiation with respect to $X$, and the pressure is
\begin{equation}
\label{eq:kesspress}
p = F \, .
\end{equation}
The speed of sound squared is further
\begin{equation}
\label{eq:kessspeed}
c_s^2=\frac{d p}{d \rho}=\frac{p_X}{\rho_X} = \frac{F_X}{F_X+2 X F_{XX}} \, ,
\end{equation}
whereas the parameter of the Equation of State has the form
\begin{equation}
\label{eq:kesspar}
w=-\frac{F}{F-2 X F_X} \, .
\end{equation}

The formalism of purely kinetic k-essence was successfully combined with the $c_s^2(w)$ models and a method of constructing $F(X)$ functions for this class of models was introduced and elaborated in \cite{mi4}. In our considerations we start from two expressions which define $F(X)$ and $X$ as functions of $w$:  
\begin{equation}
\label{eq:Fofw}
\frac{d F}{F} = \frac{c_s^2}{w(c_s^2-w)} d w \, ,
\end{equation}
and
\begin{equation}
\label{eq:Xofw}
\frac{d X}{X} = 2 \frac{c_s^2}{(1+w)(c_s^2-w)} d w \, .
\end{equation}
The details of derivations of these expressions can be found in \cite{mi4}. A final equation determining the scale factor dependence of $w$ in $c_s^2$ models is
\begin{equation}
\label{eq:wofa_dyn}
\frac{d w}{(c_s^2-w)(1+w)} = -3 \frac{d a}{a} \, .
\end{equation}

A crucial assumption of the new reconstruction method is to introduce a new function $f(w)$  such that  
\begin{equation}
\label{eq:deff}
\frac{c_s^2}{(1+w)(c_s^2-w)} = f'(w) \, .
\end{equation}
{\bf Here prime in $f'$ refers to a derivative of the function $f$ with respect to its argument.} This relation serves as an implicit definition of $c_s^2$ as a function of $w$. An explicit expression for $c_s^2(w)$ is then
\begin{equation}
\label{eq:cs2ofw}
c_s^2=-\frac{w (1+w)f'(w)}{1-(1+w) f'(w)} \, .
\end{equation}
Furher it follows
\begin{equation}
\label{eq:Xoff}
\ln \frac{X}{X_0} = 2(f(w)-f(w_0)) \, ,
\end{equation}
and, assuming that $f$ is invertible,
\begin{equation}
\label{eq:wofX}
w=f^{-1}\left(f(w_0)+\frac{1}{2} \ln \frac{X}{X_0} \right) \, .
\end{equation}
On the other hand,
\begin{equation}
\label{eq:Foff}
\ln \frac{F}{F_0} = \frac{1}{2} \ln \frac{X}{X_0} + \int_{w_0}^{w} \frac{f'(w')}{w'} d\, w' \, .
\end{equation}
Furthermore, the dynamics of $w(a)$ can be expressed in closed form as
\begin{equation}
\label{eq:wofa}
\frac{w_0}{w} \frac{w+1}{w_0+1} e^{\int_{w_0}^w \frac{f'(w')}{w'} d\, w'} =\left(\frac{a}{a_0} \right)^{-3} \, .     
\end{equation}
{\bf In the expressions (\ref{eq:Foff}) and (\ref{eq:wofa}) $w'$ is an integration variable.} Therefore, to have analytically tractable models (in the sense that analytic expressions for $F(X)$ can be obtained and that a closed expression for $w(a)$ can be obtained), two conditions must be satisfied:
\begin{itemize}
\item $f(w)$ should be invertible,
\item $f'(w)/w$ should be integrable.
\end{itemize}
Invertibility of $f(w)$ is in some cases secured by the restriction of $w$ to a finite interval (e.g. a $[-1,0]$ interval in parametric regimes that correspond to unified dark sector descriptions).  

{\bf Although the reconstruction method presented in this paper is based on the formalism developed in \cite{mi4}, it also presents a novel step in finding analytical solutions for the $c_s^2(w)$ models. That is, both \cite{mi4} and this paper start from expressions (\ref{eq:Fofw}) and (\ref{eq:Xofw}). In the approach chosen in \cite{mi4}, an explicit form of $c_s^2(w)$ is chosen, and (\ref{eq:Fofw}) and (\ref{eq:Xofw}) are integrated and inverted to obtain the expression for $F(X)$. On the other hand, in this paper we start from a $f(w)$ function and proceed as specified in the expressions presented above. The approaches presented in \cite{mi4} and in this paper are complementary, in the sense that they are suitable for the analytical reconstruction of different models. More precisely, some models which are analytically tractable using the approach of \cite{mi4} may not be analytically tractable using the method from this paper and vice versa. An additional novelty of the method introduced in this paper is the generality of its applicability, summarized in invertibility of $f(w)$ and integrability of $f'(w)/w$.}

{\bf As the starting point of the method presented here is a function $f(w)$ (or more precisely its derivative), an important question is which choices of this function are physically interesting, especially since from (\ref{eq:deff}) we see that $f(w)$ is not a simple function of physically relevant quantities. Our main motivation in this paper is to demonstrate the usefulness of the introduced method and we, therefore, select those $f(w)$ functions for which the entire procedure can be carried out analytically. However, from (\ref{eq:cs2ofw}) it is clear that $c_s^2(w)$ is a nontrivial function which allows consideration of various scenarios of cosmic evolution, as discussed later in the paper.}

In the next section we demonstrate the method introduced in this section and provide analytical solutions for five distinct models.

\section{Analytically tractable models}

The models considered in this section have been primarily chosen as examples where the method introduced in the preceding section can be carried out analytically. Solutions are provided as explicit (or at least closed-form) expressions. For all considered models, analytical solutions are accompanied by plots illustrating parametric regimes useful for the understanding of the cosmic dark sector.

\subsection{Model 1}

\begin{figure}[!t]
\centering
\includegraphics[scale=0.7]{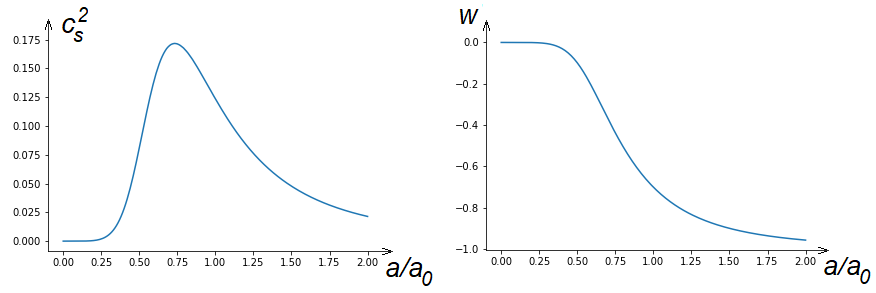}
\caption{Graphs of $c_s^2(a)$ and $w(a)$ functions for Model 1 and parameter values $w_0 = -0.7$ and $C = 0.5$. A distinguished feature is a peak in $c_s^2(a)$ at $a/a_0=0.73$.}
\label{fig_Model1}
\end{figure}


Let us further consider a concrete example:

\begin{equation}
\label{eq:model1fdef}
f'(w)= C \, ,
\end{equation}
where $C$ is a constant, such that $0 < C < 1$. {\bf Here, and in all other models considered in this paper, subscript 0 in quantities such as $w$, $X$, $F$ or scale factor $a$ refers to a present value of specific quantity (e.g. $w_0$ is a present value of the quantity $w$).} Then it follows
\begin{equation}
\label{eq:model1fofw}
f(w)-f(w_0)=C(w-w_0)\, ,
\end{equation}
which, using (\ref{eq:Xoff}), leads to 
\begin{equation}
\label{eq:model1wofX}
w=w_0 + \frac{1}{2 C} \ln \frac{X}{X_0} \, .
\end{equation}
Further 
\begin{equation}
\label{eq:model1integral}
\int_{w_0}^{w} \frac{f'(w')}{w'} d\, w' = C \ln \frac{w}{w_0} \, ,
\end{equation}
which finally leads to
\begin{equation}
\label{eq:model1Ffinal}
F=F_0 \left(\frac{X}{X_0} \right)^{1/2} \left[\ 1 + \frac{1}{2 Cw_0} \ln \frac{X}{X_0} \right]^C \, .
\end{equation}
The speed of sound squared is then from (\ref{eq:model2fdef})
\begin{equation}
\label{eq:model1cs2ofw}
c_s^2=-\frac{ C (1+w) w}{1- C (1+w) } \, ,
\end{equation}
and the dynamics of $w(a)$ is implicitly defined by
\begin{equation}
\label{eq:model1wofa}
\left (\frac{w}{w_0} \right)^{C-1} \frac{w+1}{w_0+1}  = \left(\frac{a}{a_0} \right)^{-3}\, .
\end{equation}

\subsection{Model 2}

\begin{figure}[!t]
\centering
\includegraphics[scale=0.7]{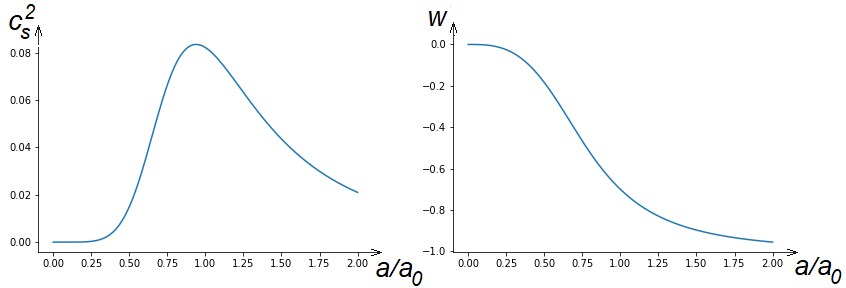}
\caption{Graphs of $c_s^2(a)$ and $w(a)$ functions for Model 2 and parameter values $w_0 = -0.7$, $\mu = 0.5$ and $\nu = 1$. A distinguished feature is a peak in $c_s^2(a)$ at $a/a_0=0.94$. }
\label{fig_Model2}
\end{figure}


Next we consider the following model:

\begin{equation}
\label{eq:model2fdef}
f'(w)=\mu (-w)^{\nu} \, ,
\end{equation}
{\bf where $\mu$ and $\nu$ are real constants}. From (\ref{eq:model2fdef}) one obtains
\begin{equation}
\label{eq:model2fofw}
f(w)-f(w_0)=-\frac{\mu}{\nu+1}\left[(-w)^{\nu+1} - (-w_0)^{\nu+1} \right]\, ,
\end{equation}
which, combined with (\ref{eq:Xoff}), results in 
\begin{equation}
\label{eq:model2wofX}
w=-\left[(-w_0)^{\nu+1} - \frac{\nu+1}{2 \mu} \ln \frac{X}{X_0} \right]^{\frac{1}{\nu+1}}\, .
\end{equation}
The required integral is then 
\begin{equation}
\label{eq:model2integral}
\int_{w_0}^{w} \frac{f'(w')}{w'} d\, w' =\frac{\mu}{\nu} \left[(-w)^{\nu} - (-w_0)^{\nu} \right]\, ,
\end{equation}
which results in the explicit expression for $F(X)$, i.e.
\begin{equation}
\label{eq:model2Ffinal}
F=F_0 \left(\frac{X}{X_0} \right)^{1/2} e^{\frac{\mu}{\nu} \left[\left[(-w_0)^{\nu+1} - \frac{\nu+1}{2 \mu} \ln \frac{X}{X_0} \right]^{\frac{\nu}{\nu+1}} - (-w_0)^{\nu} \right]} \, .
\end{equation}
The speed of sound squared follows from (\ref{eq:model2fdef})
\begin{equation}
\label{eq:model2cs2ofw}
c_s^2=\mu \frac{(1+w) (-w)^{\nu+1}}{1-\mu (1+w) (-w)^{\nu}} \, ,
\end{equation}
whereas the dependence of $w $ on $a$ is given in an implicit form by
\begin{equation}
\label{eq:model2wofa}
\frac{w_0}{w} \frac{w+1}{w_0+1} e^{\frac{\mu}{\nu} \left[(-w)^{\nu} - (-w_0)^{\nu} \right] } = \left(\frac{a}{a_0} \right)^{-3}\, .
\end{equation}

\subsection{Model 3}

\begin{figure}[!t]
\centering
\includegraphics[scale=0.67]{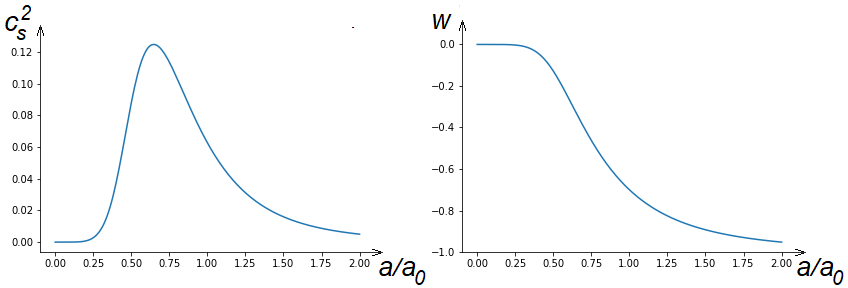}
\caption{Graphs of $c_s^2(a)$ and $w(a)$ functions for Model 3 and parameter values $w_0 = -0.7$ and $A = 0.5$. A distinguished feature is a peak in $c_s^2(a)$ at $a/a_0=0.65$.}
\label{fig_Model3}
\end{figure}


The following model that we consider is defined by

\begin{equation}
\label{eq:model3fdef}
f'(w)=A \sqrt{1+w} \, ,
\end{equation}
where $A$ is a constant. This model is well defined only if $w \ge -1$ throughout the entire cosmic timeline. 
It is further straightforward to obtain
\begin{equation}
\label{eq:model3fofw}
f(w)-f(w_0)=\frac{2 A}{3} \left[(1+w)^{\frac{3}{2}} - (1+w_0)^{\frac{3}{2}} \right]\, ,
\end{equation}
which, using (\ref{eq:Xoff}), gives the expression for $w$ as a function of $X$ 
\begin{equation}
\label{eq:model3wofX}
w=-1 + \left[(1+w_0)^{\frac{3}{2}} + \frac{3}{4 A} \ln \frac{X}{X_0} \right]^{\frac{2}{3}}\, .
\end{equation}
The integral needed for the reconstruction of the $F(X)$ function is then 
\begin{equation}
\label{eq:model3integral}
\int_{w_0}^{w} \frac{f'(w')}{w'} d\, w' = 2 A (\sqrt{1+w}-\sqrt{1+w_0}) + A \ln \left[\frac{(\sqrt{1+w}-1)(\sqrt{1+w_0}+1)}{(\sqrt{1+w}+1)(\sqrt{1+w_0}-1)} \right]\, ,
\end{equation}
which results in the expression
\begin{eqnarray}
\label{eq:model3Ffinal}
F &=&F_0 \left(\frac{X}{X_0} \right)^{1/2} \left[\frac{\sqrt{1+w_0}+1}{\sqrt{1+w_0}-1} \frac{\left[(1+w_0)^{\frac{3}{2}} + \frac{3}{4 A} \ln \frac{X}{X_0} \right]^{\frac{1}{3}}-1}{\left[(1+w_0)^{\frac{3}{2}} + \frac{3}{4 A} \ln \frac{X}{X_0} \right]^{\frac{1}{3}}+1} \right]^A \\
&\times& e^{2A \left[\left[(1+w_0)^{\frac{3}{2}} + \frac{3}{4 A} \ln \frac{X}{X_0} \right]^{\frac{1}{3}} - \sqrt{1+w_0} \right]}\, .
\end{eqnarray}
Starting from (\ref{eq:model3fdef}), the speed of sound squared is 
\begin{equation}
\label{eq:model3cs2ofw}
c_s^2=-\frac{A w (1+w)^{\frac{3}{2}}}{1-A (1+w)^{\frac{3}{2}}} \, ,
\end{equation}
and the implicitly defined function $w(a)$ is then
\begin{equation}
\label{eq:model3wofa}
\frac{w_0}{w} \frac{w+1}{w_0+1} \left[\frac{(\sqrt{1+w}-1)(\sqrt{1+w_0}+1)}{(\sqrt{1+w}+1)(\sqrt{1+w_0}-1)} \right]^A e^{2 A \left(\sqrt{1+w}-\sqrt{1+w_0} \right) } = \left(\frac{a}{a_0} \right)^{-3}\, .
\end{equation}

\subsection{Model 4}

\begin{figure}[!t]
\centering
\includegraphics[scale=0.67]{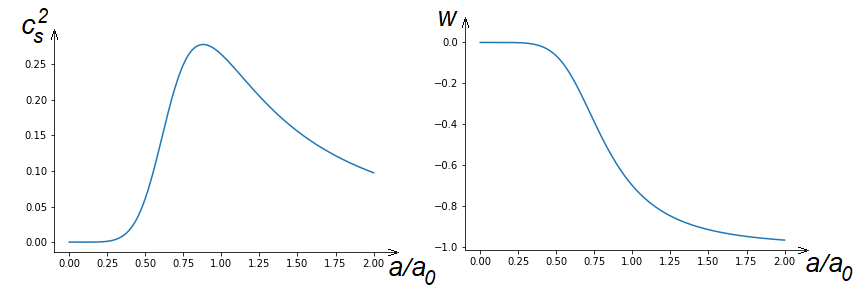}
\caption{Graphs of $c_s^2(a)$ and $w(a)$ functions for Model 4 and parameter values $w_0 = -0.7$ and $A = 0.5$. A distinguished feature is a peak in $c_s^2(a)$ at $a/a_0=0.88$.
}
\label{fig_Model4}
\end{figure}


Another analytically tractable model follows from

\begin{equation}
\label{eq:model4fdef}
f'(w)=\frac{A}{\sqrt{1+w}} \, ,
\end{equation}
where $A$ is a constant. Like in the case of Model 3, this model is well defined only if $w \ge -1$ for all scale factor values. 
The form of $f(w)$ function is then
\begin{equation}
\label{eq:model4fofw}
f(w)-f(w_0)=2 A \left[\sqrt{1+w} - \sqrt{1+w_0} \right]\, ,
\end{equation}
which, using (\ref{eq:Xoff}), results in 
\begin{equation}
\label{eq:model4wofX}
w=-1 + \left[\sqrt{1+w_0} + \frac{1}{4 A} \ln \frac{X}{X_0} \right]^2\, .
\end{equation}
The integration of $f(w)/w$ gives 
\begin{equation}
\label{eq:model4integral}
\int_{w_0}^{w} \frac{f'(w')}{w'} d\, w' =  A \ln \left[\frac{(\sqrt{1+w}-1)(\sqrt{1+w_0}+1)}{(\sqrt{1+w}+1)(\sqrt{1+w_0}-1)} \right]\, ,
\end{equation}
leading to
\begin{equation}
\label{eq:mode4Ffinal}
F = F_0 \left(\frac{X}{X_0} \right)^{1/2} \left[ \frac{1 + \frac{1}{4 A(\sqrt{1+w_0}-1)} \ln \frac{X}{X_0} }{1 + \frac{1}{4 A (\sqrt{1+w_0}+1)} \ln \frac{X}{X_0}} \right]^A \, .
\end{equation}
The speed of sound squared is then obtained from (\ref{eq:model4fdef})
\begin{equation}
\label{eq:model4cs2ofw}
c_s^2=-\frac{A w \sqrt{1+w}}{1-A \sqrt{1+w}} \, ,
\end{equation}
and the dynamics of $w(a)$ acquires the following implicit form
\begin{equation}
\label{eq:model4wofa}
\frac{w_0}{w} \frac{w+1}{w_0+1} \left[\frac{(\sqrt{1+w}-1)(\sqrt{1+w_0}+1)}{(\sqrt{1+w}+1)(\sqrt{1+w_0}-1)} \right]^A  = \left(\frac{a}{a_0} \right)^{-3}\, .
\end{equation}

\subsection{Model 5}

\begin{figure}[!t]
\centering
\includegraphics[scale=0.67]{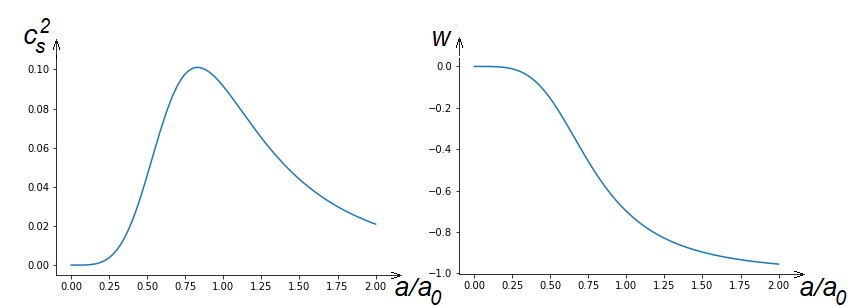}
\caption{Graphs of $c_s^2(a)$ and $w(a)$ functions for Model 5 and parameter values $w_0 = -0.7$, $A = 0.5$ and $w_* = -2$. A distinguished feature is a peak in $c_s^2(a)$ at $a/a_0=0.83$. }
\label{fig_Model5}
\end{figure}


Finally we consider a model defined by

\begin{equation}
\label{eq:model5fdef}
f'(w)=\frac{A}{w-w_*} \, ,
\end{equation}
where $A$ and $w_*$ are constants. Let us further assume that $w_* \neq -1$ and $A \neq w_*$. One  obtains
\begin{equation}
\label{eq:model5fofw}
f(w)-f(w_0)=A \ln \frac{w-w_*}{w_0-w_*} \, ,
\end{equation}
which, using (\ref{eq:Xoff}), results in 
\begin{equation}
\label{eq:model5wofX}
w= w_* + (w_0-w_*)\left( \frac{X}{X_0} \right)^{\frac{1}{2A}}\, .
\end{equation}
Further it is straightforward to obtain
\begin{equation}
\label{eq:model5integral}
\int_{w_0}^{w} \frac{f'(w')}{w'} d\, w' = \frac{A}{w_*} \left[ \ln \frac{w-w_*}{w_0-w_*} - \ln \frac{w}{w_0}\right]    \, ,
\end{equation}
which finally gives an expression
\begin{equation}
\label{eq:model5Ffinal}
F=F_0 \frac{\left(\frac{X}{X_0} \right)^{\frac{1}{2} \left(1+\frac{1}{w_*} \right)}}{\left[ \frac{w_*}{w_0} + \left(1 - \frac{w_*}{w_0} \right) \left(\frac{X}{X_0} \right)^{\frac{1}{2A}} \right]^{\frac{A}{w_*}} } \, .
\end{equation}
Using (\ref{eq:model2fdef}), the speed of sound squared is 
\begin{equation}
\label{eq:model5cs2ofw}
c_s^2=-\frac{A w (1+w)}{w-w_* -A (1+w)} \, ,
\end{equation}
and the implicitly defined function $w(a)$ is
\begin{equation}
\label{eq:model5wofa}
\left(\frac{w}{w_0} \right)^{-1-\frac{A}{w_*}} \frac{w+1}{w_0+1} \left( \frac{w-w_*}{w_0-w_*} \right)^{\frac{A}{w_*}} = \left(\frac{a}{a_0} \right)^{-3}\, .
\end{equation}
The evolution of $w(a)$ depends on the concrete value of the $w_*$ parameter. For $-1 < w_0 < 0$, and a sufficiently small positive $A$, if $w_* < -1$, the $w(a)$ function interpolates between 0 and -1. If $-1 < w_* < w_0$, $w(a)$ is contained between 0 and $w_*$. 





\section{Discussion and conclusions}

\begin{figure}[!t]
\centering
\includegraphics[scale=0.9]{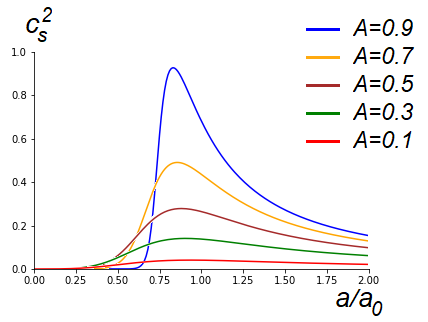}
\caption{Dependence of $c_s^2(a)$ in Model 4 for parameter values $w_0 = -0.7$ and various values of parameter $A$.}
\label{fig_Model4_varA}
\end{figure}

\begin{figure}[!t]
\centering
\includegraphics[scale=0.9]{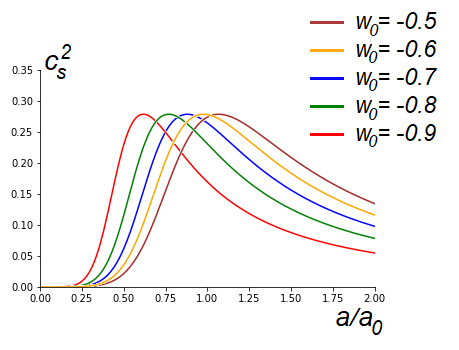}
\caption{Dependence of $c_s^2(a)$ in Model 4 for parameter values $A = 0.5$ and various values of parameter $w_0$. }
\label{fig_Model4_varw0}
\end{figure}

The focus of this paper is on the introduced method of purely kinetic k-essence model reconstruction for $c_s^2(w)$ models. The said reconstruction is explicitly analytically performed for five $c_s^2(w)$ models. The introduced method is, therefore, an alternative to approach where $c_s^2$ is modeled directly as a function of $w$, which was elaborated in \cite{mi4}.  

The considered models (models 1, 2, 3, 4 and 5) were chosen on the basis of analytical tractability of required calculations. However, for all of them there are parameter regimes where they represent a cosmic component corresponding to dark matter - dark energy unification. The dependence of $c_s^2$ and $w$ on the scale factor $a$ are presented in Fig. \ref{fig_Model1} for Model 1,  Fig. \ref{fig_Model2} for Model 2, Fig. \ref{fig_Model3} for Model 3, Fig. \ref{fig_Model4} for Model 4 and Fig. \ref{fig_Model5} for Model 5. In all figures $w$ vanishes for small $a$ and approaches -1 for large values of $a$, whereas $c_s^2$ vanishes for small and large values of the scale factor and reaches maximum at some intermediate scale factor value. 
{\bf As an illustration, in Figures \ref{fig_Model4_varA} and \ref{fig_Model4_varw0}, for the Model 4, we present the form of the $c_s^2(a)$ function for different values of parameters $A$ and $w_0$. It can be seen that the variation of parameter $A$ primarily causes the change of the height of the peak, while the location of the peak, represented  by the scale factor value when $c_s^2(a)$ reaches maximum, remains mostly unchanged. Conversely, a change of the current value of the $w$ parameter, $w_0$, keeps the height of the maximum constant, but shifts its position in the history of the universe. These results demonstrate the sensitivity of $c_s^2(a)$ on values of model parameters which is important for efficiently constraining model parameters from the observational data.}
The deviation of the speed of sound squared from 0 is controlled by one of model parameters which suggests that models might be consistent with growth of structure data for sufficiently small values of corresponding parameters. This claim can be fully statistically verified only by the model comparison with the real cosmological observations, which is out of scope of this paper. {\bf However, some qualitative arguments can be made on the potential physical importance of the studied models. As the speed of sound squared exhibits variation with the cosmic expansion, the growth of inhomogeneities and formation of cosmic structure in these models could be enhanced or suppressed at various scales and in different phases of cosmic expansion. This impact on structure formation is more prominent at redshifts when $c_s^2(a)$ peaks. This fact potentially opens new pathways towards resolution of presents tensions in cosmology, such as $H_0$ and $S_8$ tensions.} Furthermore, all parametric regimes of the constructed models have not been thoroughly examined because it would exceed the scope of this paper. It is certain that there is a number of additional parameter regimes, which were not presented here, that may be of interest. {\bf A question of particular importance is if in these regimes $c_s^2(w)$ could in some phases become negative, which then leads to related questions of stability of the unified dark sector. Potential instabilities of this kind could provide strong constraints on model parameters in comparison against observational data.} In conclusion, although the considered models were chosen to demonstrate the introduced reconstruction method, they are, in adequate parameter regimes, also interesting as possible explanations of the cosmic dark sector.  

Availability and straightforwardness of solutions to the above models depends crucially on the integrability and invertibility of functions used in the reconstruction method. With a careful approach, it is possible to find a number of functions $f(w)$ that can satisfy these conditions.

As already discussed in \cite{mi4}, even if $f(w)$ function is not analytically invertible, one may obtain parametric solutions $(X(w),F(w))$, which can be used to obtain parametric plots of $F(X)$. Furthermore, since there are parameter regimes in which $w$ is constrained to a $[-1,0]$ interval, a numerical analysis of such parametric solutions is simplified. {\bf That is, even for the most complex models for which the analytical integration of the r.h.s. of (\ref{eq:Fofw}) and (\ref{eq:Xofw}) is not feasible, it is possible to perform numerical integration of these expression with an advantage that numerical integration need not be performed on an infinite interval, but on a finite interval since in many physically interesting regimes $w$ is confined to the $[-1,0]$ interval. This advantage is also applicable to the method presented in this paper, i.e. in more general models, which are not analytically tractable, required inversion in (\ref{eq:wofX}) and integrations in (\ref{eq:Foff}) and (\ref{eq:wofa}) have to be done only on the $w$ interval $[-1,0]$ for physically interesting cases.}

In conclusion, the introduced method allows analytical reconstruction of an equivalent purely kinetic k-essence description of a broad class of $c_s^2(w)$ models. The method is demonstrated on five $c_s^2(w)$ models. The analyzed models can be useful in the study of dak matter - dark energy unified dark sector in appropriate parameter regimes.






\begin{thebibliography}{}

\bibitem{SNIa1} A. G. Riess et al., Astron. J. 116 (1998) 1009.

\bibitem{SNIa2} S. Perlmutter et al., Astrophys. J. 517 (1999) 565.

\bibitem{CMB1} P. A. R. Ade et al. (Planck) Astron. Astrophys. 594 (2016) A13.

\bibitem{CMB2} E. Komatsu et al, Prog. Theor. Exp. Phys. 6 (2014) 06B102.

\bibitem{BAO} C. Alcock et al., Phys. Rev. Lett. 74 (1995) 2867.

\bibitem{DERev1} D. Huterer, D. L. Shafer,  Rept. Prog. Phys. 81 (2018) 016901.

\bibitem{DERev2} P. Brax,  Rept. Prog. Phys. 81 (2018) 016902.

\bibitem{DERev3} E. J. Copeland, M. Sami, S. Tsujikawa, Int. J. Mod. Phys. D15 (2006) 1753.

\bibitem{DERev4} J. Frieman, M. Turner, D. Huterer, Ann. Rev. Astron. Astrophys. 46 (2008) 385.

\bibitem{DERev5} K. Bamba, S. Capozziello, S. Nojiri, S. D. Odintsov, Astrophys. Space Sci. 342 (2012) 155.

\bibitem{DEparam} S. K. J. Pacif, R. Myrzakulov, S. Myrzakul, Int. J. Geom. Math. Phys. 14 (2017) 1750111.

\bibitem{ModGravRev1} S. Nojiri, S.D. Odintsov, V.K. Oikonomou,  Phys. Rept. 692 (2017) 1.

\bibitem{ModGravRev2} A. De Felice, S. Tsujikawa, Living Rev.Rel. 13 (2010) 3.

\bibitem{ModGravRev3} T. P. Sotiriou, V. Faraoni, Rev. Mod. Phys. 82 (2010) 451.

\bibitem{ModGravRev4} S. D. Odintsov, V. K. Oikonomou, I. Giannakoudi, F. P. Fronimos, E. C. Lymperiadou, Symmetry 15 (2023) 9.

\bibitem{DM1} S. Tulin, H.-B. Yu, Phys. Rept. 730 (2018) 1.

\bibitem{DM2} M. Klasen, M. Pohl, G. Sigl, Prog. Part. Nucl. Phys. 85 (2015) 1.

\bibitem{DM3} G. Bertone, D. Hooper, Rev. Mod. Phys. 90 (2018)  045002.

\bibitem{DM4} R. Barkana, Nature 555 (2018) 71.

\bibitem{MOND1} M. Milgrom, Astrophys. J 270 (1983) 371.

\bibitem{MOND2} D. V. Bugg, Can. J. Phys. 93 (2015) 119.

\bibitem{MOND3} S. McGaugh, F. Lelli, J. Schombert, Phys. Rev. Lett. 117 (2016) 201101.

\bibitem{MOND4} C. Skordis, T. Zlosnik, Phys. Rev. Lett. 127 (2021) 161302.
    
\bibitem{MOND5} S. Vagnozzi, Class. Quant. Grav. 34 (2017) 185006.
    
\bibitem{H0tension1} E. Di Valentino et al., Astropart. Phys. 131 (2021) 102605. 
    
\bibitem{H0tension2} E. Di Valentino, O. Mena, S. Pan, L. Visinelli, W. Yang, A. Melchiorri, D. F. Mota, A. G. Riess, J. Silk, Class. Quantum Grav. 38 (2021) 153001.
    
\bibitem{H0tension3} E. Abdalla et al.,  J. High En. Astrophys. 2204 (2022) 002.  
    
\bibitem{H0tension4} S. Vagnozzi, Phys. Rev. D 102 (2020) 023518.

\bibitem{H0tension5} S. Vagnozzi, Universe 9 (2023) 393.

\bibitem{Bertacca} D. Bertacca, N. Bartolo, S. Matarrese, Adv. Astron. 2010 (2010) 904379.

\bibitem{LinderScherrer} E. V. Linder, R. J. Scherrer, Phys. Rev. D 80 (2009) 023008.

\bibitem{Chap1} A.Y. Kamenshchik, U. Moschella, V. Pasquier, Phys. Lett. B 511 (2001) 265.

\bibitem{Chap2} N. Bilic, G. B. Tupper, R. D. Viollier, Phys. Lett. B 535 (2002) 17.

\bibitem{GenChap1} M. C. Bento, O. Bertolami, A. A. Sen, Phys. Rev. D 66 (2002), 043507.

\bibitem{ModChap1} J. D. Barrow, Nucl. Phys. B 310 (1988) 743.

\bibitem{ModChap2} H. B. Benaoum, arXiv:hep-th/0205140v1.

\bibitem{HybridChap} N.  Bilic,  G.  B.  Tupper,  R.  D.  Viollier,  JCAP 0510 (2005)  003.

\bibitem{Umami} R. Lazkoz, M. Ortiz-Ba\~{n}os, V. Salzano, Phys. Dark Univ. 24 (2019) 100279.

\bibitem{GCGconstraint} H. Sandvik, M. Tegmark, M. Zaldarriaga, I. Waga, Phys. Rev. D 69 (2004) 123524. 

\bibitem{underdetermined} W. J. Wolf, P. G. Ferreira, Phys. Rev. D 108 (2023) 103519.



\bibitem{Caplar} N. Caplar, H. Stefancic, Phys. Rev. D 87 (2013) 023510.

\bibitem{mi2} D. Perkovic, H. Stefancic, Phys. Lett. B 797 (2019) 134806.

\bibitem{mi4} D. Perkovic, H. Stefancic, Phys. Dark Univ. 32 (2021) 100827.

\bibitem{mi1} D. Perkovic, H. Stefancic, Int. J. Mod. Phys. D 28 (2018) 1950045.

\bibitem{mi3} D. Perkovic, H. Stefancic, Eur. Phys. J. C 80 (2020) 629.

\bibitem{mi5} D. Perkovic, H. Stefancic, Eur. Phys. J. C 83 (2023) 306.

\bibitem{CAP1} C. Armendariz-Picon, T. Damour, V. F. Mukhanov, Phys. Lett. B 458 (1999) 209.

\bibitem{Garriga} J. Garriga, V. F. Mukhanov, Phys. Lett. B 458 (1999) 219.

\bibitem{Chiba} T. Chiba, T. Okabe, M. Yamaguchi, Phys. Rev. D 62 (2000) 023511.

\bibitem{CAP2} C. Armendariz-Picon, V. F. Mukhanov, P. J. Steinhardt, Phys. Rev. Lett. 85 (2000) 4438.

\bibitem{CAP3} C. Armendariz-Picon, V. F. Mukhanov, P. J. Steinhardt, Phys. Rev. D 63 (2001) 103510.

\bibitem{Malquarti1} M Malquarti, E. J. Copeland, A. R. Liddle, M. Trodden, Phys. Rev. D 67 (2003) 123503.

\bibitem{Malquarti2} M. Malquarti, E. J. Copeland, A. R. Liddle, Phys. Rev. D 68 (2003) 023512.

\bibitem{Chimento} L. P. Chimento, A. Feinstein, Mod. Phys. Lett. A 19 (2004) 761.

\bibitem{dePutter} R. de Putter, E. V. Linder, Astropart. Phys. 28 (2007) 263.

\bibitem{Scherrer} R. J. Scherrer, Phys. Rev. Lett. 93 (2004) 011301. 

\bibitem{Chimento2} L. P. Chimento, M. I. Forte, R. Lazkoz, Mod. Phys. Lett.A 20 (2005) 2075.

\bibitem{Armendariz} C. Armendariz-Picon, E. A. Lim, JCAP 08 (2005) 007.

\bibitem{kessence2} D. Bertacca, S. Matarrese, M. Pietroni, Mod. Phys. Lett. A 22 (2007) 2893. 

\bibitem{Bertacca2} D. Bertacca, N. Bartolo, A. Diaferio, S. Matarrese, JCAP 10 (2008) 023.

\bibitem{kessence3} N. Bilic, G. B. Tupper, R. D. Viollier, Phys. Rev. D 80 (2009) 023515.

\bibitem{Bertacca3} O. F. Piattella, D. Bertacca, M. Bruni, D. Pietrobon, JCAP 01 (2010) 014.

\bibitem{Chavanis} P.-H. Chavanis, Astronomy 1 (2022) 126.

\bibitem{tachyon1} A. Sen, JHEP 04 (2002) 048.

\bibitem{tachyon2} A. Sen, Mod. Phys. Lett. A 17 (2002) 1797.

\bibitem{tachyon3} G. W. Gibbons, Phys. Lett. B 537 (2002) 1.

\bibitem{tachyon4} J. S. Bagla, H. K. Jassal, T. Padmanabhan, Phys. Rev. D 67 (2003) 063504.

\bibitem{kesstension} S. A. Hosseini Mansoori, H. Moshafi, Astrophys. J. 975 (2024) 275.

\bibitem{Diez} A. Diez-Tejedor, A. Feinstein, Int. J. Mod. Phys. D 14 (2005) 1561.

\bibitem{Arroja} F. Arroja, M. Sasaki, Phys. Rev. D 81 (2010) 107301.

\bibitem{Ferreira} V. M. C. Ferreira, P. P. Avelino, R. P. L. Azevedo, Phys. Rev. D 102 (2020) 063525.

\bibitem{Unnikrishnan} S. Unnikrishnan, L. Sriramkumar, Phys.Rev. D 81 (2010) 103511.

\bibitem{kessphase} I. Quiros, T Gonzalez, U. Nucamendi, R. De Arcia, F. A. Horta Rangel, arXiv:2501.14177.

\end{thebibliography}
\end{document}